\documentclass [10pt]{article}
\usepackage[T1]{fontenc}
\usepackage[final]{epsfig}
\usepackage{amssymb}
\usepackage{multicol}
\usepackage{graphics}
\usepackage{color}
\usepackage{ntimes}

\newcommand{\BE}{\begin{equation}}
\newcommand{\EE}{\end{equation}}

\begin{document}

\title{Parking in the city: an example of limited resource sharing}
\author{\textbf{Petr \v Seba${}^{1,2,3}$}\\
\small{${}^1$ University of Hradec Kr\'alov\'e, Hradec Kr\'alov\'e
- Czech Republic}\\
\small{${}^2$ Institute of Physics, Academy of Sciences of the
Czech Republic, Prague - Czech Republic}\\
\small{${}^3$ Doppler Institute for Mathematical
Physics and Applied Mathematics,}\\ \small{Faculty of Nuclear
Sciences and Physical Engineering,
Czech Technical University, Prague - Czech Republic}}

\normalsize

\maketitle

\begin{abstract}

During the attempt to park a car in the city the drivers have to
share limited resources (the available roadside). We show that this
fact leads to a predictable distribution of the distances between
the cars that depends on the length of the street segment used for
the collective  parking. We demonstrate in addition that the
individual parking maneuver is guided by generic psychophysical
perceptual correlates. Both predictions are compared with the actual
parking data collected in the city of Hradec Kralove (Czech
Republic).
\end{abstract}

\section{Introduction}
Everyone knows that to park a car in the city center is problematic.
The amount of the available places is limited and has to be shared
between an increasing number of cars. There are several attempts to
tackle this challenge by introducing parking charges, building
underground garages, implementing parking zones, advertising for
public transport and many others. The problem seems however to
persist. Deeper understanding of the related processes  is therefore
of a common interest.

There have been several mathematical attempts to tackle the parallel
parking process. The classical way to do so  is the "random car
parking model" introduced by Renyi \cite{Renyi} - see \cite{evans},
\cite{ca} for review. In this model the cars park on randomly chosen
places and once parked the cars do not leave the street. All cars
are usually assumed to be of the same length $l_0$ and the process
continues as long as all available free places ale smaller then
$l_0$. The model leads to predictions that can be easily verified.
First of all it gives a relation between the mean bumper - to -
bumper distance $\overline{D}$ and the car length: $\overline{D}\sim
0.337 \ l_0$. Further: the probability density
 $p(D)$ of the car distances $D$
behaves like \cite{ar},\cite{or}, \cite{yang}, \cite{mac}
$p(D)\approx -ln(D)$. This means that small distances between cars
are preferred.

To test this results real parking data were collected recently  in
the center of London \cite{rawal}. The average distance between the
parked cars was  $152$  cm which fits nicely with the relation
$\overline{D}\sim 0.337 \ l_0$ for $l_0=450$ cm. The predicted
probability density was however incompatible with the observed
facts. The model leads to  $p(D)\to \infty$ for $D\to 0$ whereby the
data from London display  $p(D)\to 0$ as $D\to 0$. The same behavior
has been found also in other cities \cite{seba}.

This results show that the parking process is not so simple as
assumed by the "random car parking model". First of all: although
the total available parking place remains unchanged the cars are
reshuffled many times during the day since some parked cars leave
and new cars park on the vacant places. Moreover the parking
maneuver is not trivial. It is not just a simple positioning of the
car to the parking lot.

Our aim here is to give and alternative approach to the parking
process. We will show that it can be understood as a statistical
partition of the limited parking space between the competing persons
trying to park. The partition is described by the Dirichlet
distribution with a parameter $g$. We will also show that this
parameter is in fact fixed by the capability of the driver to
exploit small distances during the parking maneuver.

 Let us focus on the
spacing distribution (bumper to bumper distances) between cars
 parked parallel to the curb.
We will assume that the street segment used for parking starts and
ends with some clear and nondisplaceable part unsuitable for
parking. It can be a driveway or turning to a side street. Otherwise
the parking segment is free of any kind of parking obstructions. We
will assume that it has a length $L$. Moreover there are not marked
parking lots or park meters inside it. So the drivers are free to
park the car anywhere in the segment provided they find an empty
space to do it. We suppose also that all cars have the same length
$l_0$.

Many cars are cruising for parking in this part of the city. So
there are not free parking lots and a car can park only when another
parked car leaves. To simplify the further formulation of the
problem and to avoid troubling with the boundary effects we assume
that the street segment under consideration form a circle. The car
spacing distribution is obtained as a steady solution of the
repeated car parking and car leaving process.

Due to the parking maneuver one needs a lot of a length $\approx 1.3
l_0$ to park. Hence in a segment of length $L$ the number of the
parked cars equals to $N\approx[L/(1.3 l_0)]$. Denoting by $D_k$ the
spacing between the car $k$ and $k+1$ we get $ \sum_{k=1}^{N}D_k = L
- N l_0$ and after a simple rescaling finally

\begin{equation}\label{total}
 \sum_{k=1}^{N}D_k = 1.
\end{equation}

Since all parking lots are occupied  the number of parked cars is
supposed to be fixed. The repeated car parking and car leaving
reshuffles however the distances $D_k$. We will treat them as
independent random variables constrained by the simplex
(\ref{total}). The distance reshuffling goes as follows: In the
first step one randomly chosen car leaves the street and the two
adjoining lots merge into a single one. In the second step a new car
parks into this empty space and splits it again into two smaller
lots. Such fragmentation and coagulation processes were discussed
intensively since they apply for instance to the computer memory
allocation  - see \cite{bertoin} for review. The related equations
are simple. If a car leaves the street and the neighboring spacings
- say the spacings $D_n,D_{n+1}$ - merge into a single lot $D$ we
get
\begin{equation}\label{cdistances}
    D=D_n+D_{n+1}+l_0.
\end{equation}
When a new car parks to $D$ it splits it into
 $\tilde D_n, \tilde D_{n+1}$:
\begin{eqnarray}
 \nonumber
  \tilde D_n &=& a(D-l_0) \\
  \tilde D_{n+1} &=& (1-a)(D-l_0).
  \label{cnewdistances}
\end{eqnarray}
where $a\in (0,1)$ is a random variable with a probability density
$q(a)$. The distribution $q(a)$ describes the parking preference of
the driver. We assume that all drivers have identical preferences,
i.e. identical $q(a)$. (The meaning of the variable $a$ is
straightforward. For $a=0$ the car parks immediately in front of the
car delimiting the parking lot from the left without leaving any
empty space. For $a=1/2$ it parks exactly to the center of the lot
$D$ and for $a=1$ it stops exactly behind the car on the right.)
Combining (\ref{cdistances}) and (\ref{cnewdistances}) gives the
distance reshuffling
\begin{eqnarray}
 \nonumber
  \tilde D_n &=& a(D_n+D_{n+1}) \\
  \tilde D_{n+1} &=& (1-a)(D_n+D_{n+1}).
  \label{cnewdistances2}
\end{eqnarray}
(The car length $l_0$ drops out.) The simplex (\ref{total}) is of
course invariant under this transformation.

The mappings (\ref{cnewdistances2}) are regarded as statistically
independent for various choices of $n$. Moreover all cars are equal.
So in the steady situation the joint distance probability density
$P(D_1,...,D_{N})$ has to be exchangeable ( i.e. invariant under the
permutation of variables) and invariant with respect to
(\ref{cnewdistances2}).  Its marginals $p(D)$ (the probability
densities of the particular spacings) are identical:
\begin{equation}\label{distrib}
    p_k(D_k)=p(D_k)=\int_{D_1+..+D_N=1} P(D_1,...,D_{N})
    dD_1..dD_{k-1}dD_{k+1}..dD_{N}.
\end{equation}

A standard approach to deal with  the simplex (\ref{total}) is to
regards $D_k$  as independent random variables normalized by a sum:
\begin{equation}\label{norma}
    D_k=\frac{d_k}{\sum_{n=1}^N d_n}.
\end{equation}
Here $d_k$ are statistically independent and identically distributed
and it is preferable to work with them. Moreover: it is obvious that
the distribution of $\{D_1,..,D_N\}$ is invariant under the
transform (\ref{cnewdistances2}) merely when the distribution of
$\{d_1,..,d_N\}$ is invariant. So let us apply the relation
(\ref{cnewdistances2}) on the variables $d_n$:
\begin{eqnarray}
 \nonumber
  \tilde d_n &=& a(d_n+d_{n+1}) \\
  \tilde d_{n+1} &=& (1-a)(d_n+d_{n+1})
  \label{cnewdistances3}
\end{eqnarray}
where $a,d_n,d_{n+1}$ are independent and $d_n,d_{n+1}$ identically
distributed . Giving the distribution $q(a)$ of $a$ we  look for
distributions of $d_n$ such that the transformed variables $\tilde
d_n,\tilde d_{n+1}$  preserve the distribution of $d_n$. The effort
is to solve the equation
\begin{equation}\label{perp}
     d \triangleq a(d+d')
\end{equation}
where $d'$ is an independent copy of the variable $d$ and the symbol
$\triangleq$ means that the left and right sides of (\ref{perp})
have identical statistical properties.

Distributional equations of this type are mathematically well
studied - see for instance \cite{dev} - although not much is known
about their exact solutions. In particular it is known that for a
given distribution $q(a)$ (describing the parking habit) the
equation (\ref{perp}) has an unique solution which can be obtained
numerically. Since we are interested in explicit results we choose
$q(a)$ from a two parametric class
 of $\beta$ distributions. Then the solution of (\ref{perp})
results from the following statement \cite{duf1}:

\textbf{Statement:} Let $d_1,d_2$ and $a$ be independent random
variables with distributions: $d_1\sim \Gamma(a_1,1)$, $d_2\sim
\Gamma(a_2,1)$ and $a\sim \beta(a_1,a_2)$. Then  $a(d_1+d_2)\sim
\Gamma(a_1,1)$.

(The symbol $\sim$ means that the related random variable has the
specified probability density.   $\Gamma(g,1), \beta(g_1,g_2)$
denote the standard gamma and beta distributions respectively.)

Since in our case  the variables $d_1,d_2$ are equally distributed
we have $g_1=g_2=g$ and  $a\sim \beta(g,g)$. So the probability
density of $a$ is symmetric in this case, i.e.  the variables $a$
and $1-a$ have the same distribution. In other words the drivers are
 not biased to park more closely to a car adjacent from the behind
or from the front.

The solution of (\ref{perp}) is in this case equal to
$d\sim\Gamma(g,1)$. The relation (\ref{norma}) returns the spacings
$D_k$ and we find that the joint probability density
$P(D_1,...,D_{N})$ is nothing but a one parameter family of the
multivariate Dirichlet distributions on the simplex (\ref{total})
\cite{wilks}:
\begin{equation}\label{dirichlet}
P(D_1,...,D_{N})=\frac{\Gamma(Ng)}{\Gamma(g)^N}D_1^{g-1}
D_2^{g-1}...D_{N}^{g-1}
\end{equation}
Its marginal (\ref{distrib}) is simply $D \sim \beta(g,(N-1)g)$.
Normalizing  the mean of $D$ to $1$ we are finally left with

\begin{equation}\label{distrib2}
    p(D)=\frac{1}{N}\beta\left(g,(N-1)g,\frac{D}{N}\right).
\end{equation}

Despite of the symmetrical parking maneuver this distribution is
asymmetric. This is a consequence of the persistent parked car
exchange and can be regarded as a collective phenomenon.

The above considerations leave the parameter $g$ free. But we show
that there is in fact a natural choice of  $g$ leading to $g=3$. The
point is that the behavior of $q(a)$ for small $a$ reflects the
capability of the driver to estimate small distances.  The collision
avoidance during the parking maneuver is guided visually and this
ability is shared  equally  by all drivers. If it applies  the
behavior of $q(a)$ for small $a$ has to be generic, i.e. independent
on the particular city or parking situation. It is just fixed by the
human perception of distance.

Distance perception is a complex task and there are several cues to
do this. Some of them are monocular (linear perspective, monocular
movement parallax etc.), others oculomotor (accommodation
convergence) and finally binocular (i.e. based on stereopsis). All
of them work simultaneously and are reliable under different
conditions - see \cite{jacobs} for more details. For the parking
maneuver however the crucial information is not the distance itself
but the estimated time-to-collision between the bumper of the
parking car and its neighbors. This time has to be evaluated using
the knowledge of the distance and velocity. It has been argued in a
seminal paper by Lee \cite{lee} that the estimated time to collision
is psychophysically evaluated using a quantity named $\tau$ . It is
defined as the inverse of the relative rate of expansion of the
retinal image of the moving object. Behavioral experiments have
indicated that $\tau$ is indeed controlling actions like contacting
surfaces by flies, birds and mammals (including humans): see
\cite{weel},\cite{hop},\cite{shra}.

When the observer moves forward in the environment, the image on the
retina expands. The rate of the expansion $\tau$ conveys information
about the observer's speed and the time to collision. Psychophysical
and physiological studies have provided abundant evidence that
$\tau$ is processed by specialized neural mechanisms in the brain
\cite{far}. We take $\tau$ to be the informative variable for the
final braking - see \cite{fajen} for review.

Let $\theta$  be the instantaneous angular size of the observed
object (for instance the front of the car we are backing to during
the parking maneuver). Then the estimated time to contact is given
by

\begin{equation}\label{tau}
     \tau=\frac{\theta}{d\theta/dt}
\end{equation}

Since $\theta (t) =2 \arctan(L_0/2D(t))$ with $L_0$ being the width
of the approached object and $D(t)$ its instantaneous distance, we
get
\begin{equation}\label{tau2}
    \tau(t)=-\frac{L_0^2+4D(t)^2}{2L_0 (dD(t)/dt)}
    \arctan\left(\frac{L_0}{2D(t)}\right).
\end{equation}

For $D>>L_0$ and a constant approach speed $v=-dD/dt$ the quantity
$\tau$ simply equals to the physical arrival time: $\tau=D/v.$ For
small distances (parking maneuver), however,  $\tau\approx D^2/(v
L_0 )$ and the estimated time to contact decreases \it quadratically
\rm with the distance.

Let us return  to the equation (\ref{cnewdistances2}). For a fixed
parking lot the final stopping distance is proportional to $a$.
Assuming that the courage to exploit small distances is \it
proportional to the estimated time to contact \rm we finally get for
the probability density $q(a)$: $q(a)\approx a^2$ for small $a$.
Since $q(a) = \beta(g,g,a)$ the behavior $q(a)\approx a^2$ fix the
parameter $g$ to $g=3$ and the normalized clearance distribution
(\ref{distrib2}) reads
\begin{equation}\label{clear}
p(D)=\frac{1}{N}\beta\left(3,3(N-1),\frac{D}{N}\right)=
\left(\frac{1}{N}\right)^{3(N-1)}\frac{\Gamma(3N)}{2\Gamma(3(N-1))}D^2(N-D)^{3N-4}
\end{equation}

SO the parameter $g$ is fixed to $3$. But the distribution still
depend on the number of cars $N$ in the parking segment. It is a
consequence of the constrain (\ref{total}). For large number of
cars, $N>>1$, the constrain (\ref{total}) does not play a
substantial role and $p(D)$ equals to $\Gamma(3,1,D)$ (this is true
in the limit $N\to\infty$). In the other extreme case with $N=1$
(the parking segment is so short that it allows the parking of a
single vehicle) the distribution $p(D)$ just reflects the parking
maneuver and is equal to $q(D)$. For parking segments of
intermediate size the theory predicts a dependence of the results on
the segment length.

To verify the predictions of the model  we measured the bumper to
bumper distances between cars parked on two different streets in the
center of Hradec Kralove (Czech Republic). Both streets were located
in a place with large parking demand and usually without any free
parking lots.  In addition one of these street (street $1$)
contained driveways to courtyards. This means that the actually
available fixed parking segments were much shorter on  this street.
In the mean 3-4 cars were able to park among two subsequent
driveways and we collected 773 car spacings under this conditions.
The second street (street $2$) was free of any dividing elements.
Here we measured altogether 699 spacings.

\begin{figure}
\begin{center}
  \includegraphics[height=9cm,width=15cm]{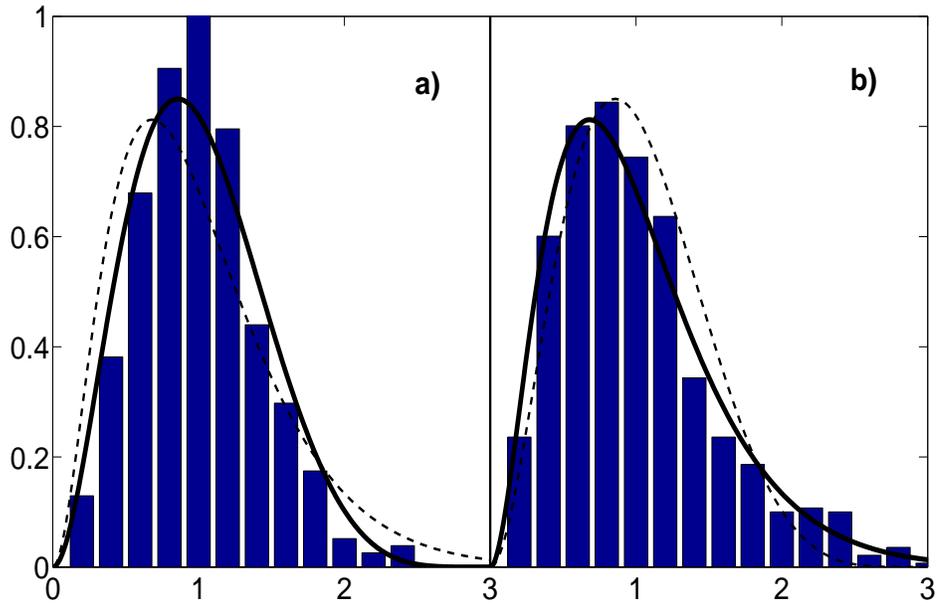}\\
\end{center}
  \caption{
The measured distance distributions for cars parking on the the
street $1$ and $2$ in the city of Hradec Kralove are compared with
the prediction of the formula (\ref{clear}). The results for the
street $1$ and $2$ are plotted on the panels a) and b) respectively.
Bars mark the probability density extracted from the collected data.
The full line stays for the formula (\ref{clear}) with $N=3$ (panel
a)) and N=20 (panel b)). To guide the eyes the result of
(\ref{clear}) for $N=20$ and $N=3$ are plotted  on the panels a) and
b) as a dashed line.
  }
  \label{parking}
\end{figure}

The probability distributions resulting from these data seems to be
fairly compatible with the prediction of the model. First of all:
the perceptual mechanism based on the estimated time to contact
seems to be verified. We demonstrated that when the estimated time
to contact is decisive for the final car stopping  then the
parameter $g$ in (\ref{distrib2}) fix to $g=3$. And exactly this
value fits with the measured data. Moreover: the finite length of
the street (the simplex (\ref{total})) leads to a dependence of the
spacing distribution (\ref{clear}) on $N$. So the result obtained
for short and long parking segments should be different. And  this
is indeed observed when the data from the street $1$ and $2$ are
compared. We plot the results on the figure \ref{parking}

The difference between the results is not large but it is
nevertheless clearly visible (compare the full and dashed lines).

To summarize we have shown that the clearance distribution for the
cars parked in paralel can be described as a marginal distribution
of the multivariate Dirichlet distribution with a parameter $g$. The
parameter is fixed to $g=3$ by the psychophysically estimated time
to collision during the parking maneuver. The measured data support
this hypothesis. The theory leads further to a prediction that the
clearance distribution depends on the length of the used parking
segment. Also this fact is verified by  the collected data.

{\bf Acknowledgement:} The research was supported by the Czech
Ministry of Education within the project  LC06002. I appreciate the
stimulating discussions with Balint Virag. The help of the PhD.
students Michal Musilek and Jan Fator, who collected the parking
data, is also gratefully acknowledged.

\end{document}